\documentclass[twocolumn,showpacs,preprintnumbers,amsmath,amssymb]{revtex4}

\usepackage{graphicx}
\usepackage{dcolumn}
\usepackage{bm}

\usepackage{color}

\usepackage{framed}
\definecolor{shadecolor}{named}{Yellow}

\begin{document}

\preprint{APS/123-QED}

\title{Shifted loops and coercivity from field imprinted high energy barriers in ferritin and ferrihydrite
nanoparticles}

\author{N. J. O. Silva}
\email{nunojoao@ua.pt}

\author{V. S. Amaral}
\affiliation{Departamento de F\'{\i}sica and CICECO, Universidade
de Aveiro, 3810-193 Aveiro, Portugal}

\author{A. Urtizberea}
\author{R. Bustamante}
\author{A. Mill\'{a}n}
\author{F. Palacio}
\affiliation{Instituto de Ciencia de Materiales de Arag\'{o}n,
CSIC - Universidad de Zaragoza. Departamento de Fisica de la
Materia Condensada, Facultad de Ciencias, 50009 Zaragoza, Spain.}
\author{E. Kampert}
\author{U. Zeitler}
\affiliation{Radboud University Nijmegen, Institut for Molecules
and Materials, High Field Magnet Laboratory , NL-6525 ED Nijmegen,
Netherlands}
\author{S. de Brion} \affiliation{Institut N\'{e}el-CNRS and Universit\'{e} Joseph Fourier BP 166-38042
Grenoble cedex 9, France}

\author{\`{O}. Iglesias}
\author{A. Labarta}
\affiliation{Departament de F\'{\i}sica Fonamental, Universitat de
Barcelona and Institut de Nanoci\`{e}ncia i Nanotecnologia,
Universitat de Barcelona, Mart\'{\i} i Franqu\`{e}s 1, 08028
Barcelona, Spain}

\date{\today}

\begin{abstract}
We show that the coercive field in ferritin and ferrihydrite
depends on the maximum magnetic field in a hysteresis loop and
that coercivity and loop shifts depend both on the maximum and
cooling fields. In the case of ferritin we show that the time
dependence of the magnetization also depends on the maximum and
previous cooling fields. This behavior is associated to changes in
the intra-particle energy barriers imprinted by these fields.
Accordingly, the dependence of the coercive and loop shift fields
with the maximum field in ferritin and ferrihydrite can be
described within the frame of a uniform-rotation model considering
a dependence of the energy barrier with the maximum and the
cooling fields.


\end{abstract}

\pacs{75.30.Cr, 75.50.Ee, 75.60.Ej, 75.50.Tt}
\maketitle

\section{\label{sec:Intro}Introduction}

The magnetic properties of ferritin have been extensively studied
in the last decade due to their puzzling features such as the
existence of a maximum in the magnetization derivative at zero
field,\cite{dMdH_prb, Tejada_prl} a nonmonotonic field dependence
of the magnetic viscosity,\cite{dMdH_prb,Tejada_prl,Mamiya_prl}
and a decrease of the antiferromagnetic susceptibility with
temperature below the N\'{e}el temperature when considered at low
fields.\cite{Berkowitz_prb,NJO_FerritinHF} Many of these studies
were performed to enlighten the possible existence of quantum
tunnelling in ferritin in the kelvin range. Less attention has
been paid to the magnetic hysteresis, although it was termed
``anomalous'' in an early report due to the large coercivity
($\sim1800$ Oe at 5 K), irreversibility up to relatively high
fields ($\sim 35$ kOe) and loop displacement found after field
cooling.\cite{Berkowitz_prb}
Ferritin consists of a hollow spherical shell composed of 24
protein subunits surrounding a ferrihydrite-like core. The
diameter of the cavity is of the order of 7-8 nm and the average
size
of the core of horse spleen ferritin is 5 nm.\cite{Mann_livro} 
The ferritin magnetic core orders antiferromagnetically 
and has both compensated and uncompensated spins, resulting in a
net magnetic moment of about a hundred of Bohr magnetons $\mu_B$
per particle.\cite{NJOS_prb}


Ferritin belongs to a group of nanoparticles where loop
displacements are found but where strictly speaking there is
neither a ferromagnetic (FM) material coupled to an
antiferromagnetic (AF) one nor a cooling across a N\'{e}el
transition temperature.
The cooling is normally performed across the blocking temperature
of the nanoparticles, i. e., across the temperature below which
the magnetic moment of the ``average'' nanoparticle is not able to
fluctuate across the anisotropy barrier within the characteristic
time of the measurement, which in ferritin and for dc measurements
is about 20 K. 
In these systems, the origin of the loop shift is not clearly
established. In AF nanoparticles the loop shift has been often
interpreted as an exchange bias between the AF core and the
uncompensated spins of the spin-glass surface. In the case of FM
and ferrimagnetic nanoparticles the loop shift is thought to arise
due to exchange between the spin-glass spins and the FM
core.\cite{Noges_Rev,Oscar_JNN} Other studies attribute the loop
shift to the existence of a minor hysteresis
loop.\cite{MinorLoopsFeCo} In fact, as highlighted in
Ref.\cite{Noges_Rev,Oscar_JNN}, many of these systems show high
field irreversibility and non-saturating hysteresis loops, raising
the question of the influence of the minor loops on the exchange
field and, eventually, the question of the existence of a real
exchange bias.

Here we report a study on the coercive field and loop shifts in
ferritin at low temperature, obtained in magnetization cycles
recorded for different maximum fields up to 30 $\times 10^4$ Oe,
and after cooling under the influence of fields with different
intensities. This study is also extended to ferrihydrite
nanoparticles, which are similar to the ferritin magnetic cores,
and complemented by the measurement of the time dependence of the
magnetization near zero field.

\section{Background}


Within the framework of the uniform rotation models developed by
Stoner-Wohlfarth and N\'{e}el-Brown,\cite{Neel_arrhenius, Brown}
reversal of magnetization over an energy barrier $E$ separating
two minima is a coherent process, which can occur at $T=0$ K when
the barrier is lowered down to zero by applying a magnetic field
in the opposite direction of the particles
magnetization,\cite{Stoner} or it can occur by thermal activation
when the thermal energy $k_BT$ becomes comparable to $E$ and thus
the characteristic reversal time $\tau$ becomes comparable to the
characteristic measurement time $\tau_m$,\cite{Neel_arrhenius}.
These quantities are related by the Arrhenius law
\begin{equation}
\label{Neel} \tau=\tau_0\exp\left(\frac{E}{k_BT}\right)\ ,
\end{equation}
where $\tau_0$ is the inverse of an attempt frequency, supposed
constant for simplicity in many situations. 
The energy barrier is field-dependent and it can be written as
\begin{equation}
\label{Energy} E(H)=E_0\left(1-\frac{H}{H_0}\right)^\alpha
\end{equation}
with $\alpha=2$ for systems with uniaxial anisotropy and easy axes parallel to the applied
field, where $E_0$ is the energy barrier at zero field
and $H_0$ is the switching field at zero temperature. In
ferromagnetic materials $H_0=2K/M_S$ and $E_0=KV$, while in general
\begin{equation}
\label{EKV} E_0=KV^p \ ,
\end{equation}
where $V$ is the particle volume, $K$ is the anisotropy constant,
$M_S$ is the saturation magnetization and $p$ an exponent equal to
1/2 in the case of antiferromagnetic ferrihydrite
nanoparticles.\cite{NJO_dist}
In the framework of the N\'{e}el model and for a random
distribution of anisotropy axes, $\alpha=4/3$
\cite{Rivas_HC_random}. In general, simulations and experimental
results show that $\alpha$ depends on the anisotropy, distribution
of particle sizes and on interparticle
interactions.\cite{Pfeiffer_pss, Nobel_jmmm} 

According to Eq. (\ref{Energy}), magnetization reversal occurs at
the coercive field $H_C$, when the energy barrier $E(H_C)$ becomes
small enough to be overcome at the given $T$ and measuring time
$\tau$
\begin{equation}
\label{HC_E}
H_C=H_0\left[1-\left(\frac{E(H_C)}{E_0}\right)^{1/\alpha}\right]\ ,
\end{equation}
where we have just re-written Eq. (\ref{Energy}). The dependence
of $H_C$ on temperature, nanoparticle volume and characteristic
measurement time can then be obtained by using Eq. (\ref{Neel})
for $E(H_C)$ in the previous equation, as shown in
Ref.\cite{Hc_vs_size_2, Hc_vs_size}. In particular, for the volume
dependence of $H_C$ at constant temperature one has
\begin{equation} \label{H_CT}
H_C(V)=H_0\left[1-(V_B/V)^{(p/\alpha)}\right]\quad
\end{equation}
where $V_B$ is the blocking volume, i.e., the volume above which
$E_0$ cannot be crossed within $\tau_m$ in a system with
anisotropy
$K$ and at a temperature $T$.  
A more refined expression for $H_C(T)$ can be obtained by
considering the temperature dependence of $K$ and
$M_S$.\cite{Effect_K(T)_H_C}

\section{Experimental}
\label{Sec:Exp}
Horse spleen ferritin samples used in these experiments were
obtained from Sigma Chemical Company and prepared in powder
samples by evaporation of the solvent at room temperature.

The synthesis of the ferrihydrite nanoparticles in the
organic-inorganic matrix (termed di-ureasil) has been described
elsewhere.\cite{NJOS_jap} The particles are precipitated by
thermal treatment at 80 $^\circ$C, after the incorporation of iron
nitrate in the matrix. The sample here studied has an iron
concentration of 2.1 \% in weight and the particles have a
diameter distribution that can be described by a lognormal
function
\begin{equation}
\label{lognormal} f(D)=\frac{1}{ D s_D \sqrt{2\pi}}
\exp\left[-\frac{(\log(D/n_D))^2}{{2s_D^2}}\right]
\end{equation}
with $n_D= 4.7\pm 0.2$ nm and standard deviation of the natural
logarithm of the diameter $s_D=0.43\pm 0.05$.\cite{NJOS_jap_tem}

For ferritin, magnetization was measured as a function of field up
to different maximum fields $H_{max}$ (in the 0.5$\times10^{4}$ to
$30\times10^{4}$ Oe range) and after cooling from 100 K down to
low temperature (3.2 and 4.2 K) in zero-field cooling (ZFC) and
field-cooling (FC) procedures using different cooling fields
$H_{cool}$. These measurements were performed in a PPMS system
(Quantum Design) with a vibrating sample magnetometer (VSM)
option, and in a Bitter magnet with a VSM (HFML facility,
Nijmegen). 
In the latter, the modulus of the magnetization was measured and
magnetization curves were reconstructed by using the proper
signal. Near zero this procedure is not perfect (since noise is
always additive) and a small kink around zero field appears (see
Fig.
\ref{Fig:M_Hmax}). 

At the characteristic time of measurement, irreversibility
phenomena vanish for $T>40$ K and magnetization curves taken at
3.2 K after cooling with $H_{cool}=2\times10^{4}$ Oe from
temperatures between 50 and 300 K are similar. The magnetization
curves are also independent on the cooling rate (cooling from 100
K) in the 0.5 to 5 K/min range. 
In addition, magnetization was measured as a function of time
during about 1000 s [$M(t)$] after cooling from 100 K down to 4.5
K in ZFC and FC with $H_{cool}=0.5\times10^{4}$ Oe procedures. For
each cooling procedure and at low temperature (4.5 K), we have
followed two different measurement protocols: i) applied different
$H_{max}$, removed the field down to a value near zero (50 Oe) and
measured $M(t)$ and ii) applied different $H_{max}$, then reversed
the field to $-H_{max}$, removed it down to $-50$ Oe and measured
$M(t)$.

For the ferrihydrite nanoparticles grown in the organic-inorganic
hybrid matrix, magnetization was measured as a function of field
up to different maximum fields $H_{max}$ (in the $2\times10^{4}$
to $20\times10^{4}$ Oe range) and after cooling from 100 K down to
3.2 K in FC procedure using $H_{cool}=2\times10^{4}$ Oe, in a
Bitter magnet with an extraction magnetometer (GHMFL facility,
Grenoble).

\section{Results and discussion}

\subsection{Effect of $H_{max}$ and $H_{cool}$ in the magnetization loops}

\begin{figure}[htb!]
\begin{center}\includegraphics[width=0.9\columnwidth]{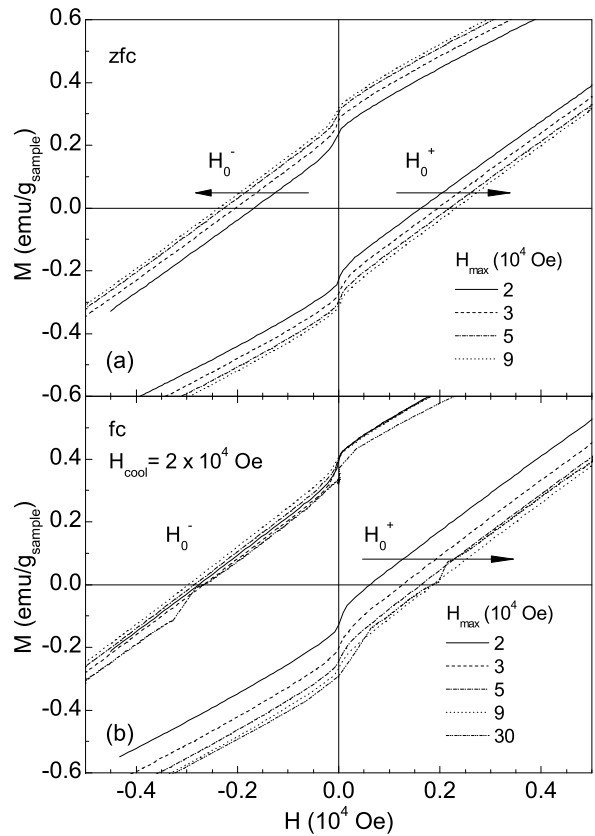}
\end{center}
\caption{\label{Fig:M_Hmax} 
Detail of magnetization loops for ferritin obtained at 3.2 K (4.2
K in the case of $H_{max}=30\times 10^4$ Oe) after ZFC (a) and FC
with $H_{cool}=2\times 10^4$ Oe (b) measured for different
$H_{max}$. $H_0^-$ and $H_0^+$ correspond to the field values at
which magnetization crosses zero in the decreasing and increasing
field branches of the hysteresis loop, respectively.}
\end{figure}

The magnetization loops of ferritin obtained at low temperature
are analogous to those previously
reported,\cite{Berkowitz_prb,dMdH_prb,ferritin_JMMM} with
non-saturation, high-field irreversibility and a maximum in the
magnetization derivative at zero field. The ZFC cycles are
symmetric and increasingly broader as $H_{max}$ increases [Fig.
\ref{Fig:M_Hmax}(a)].
When the sample is cooled in the presence of a field $H_{cool}$
the magnetization curves are shifted in the $H$ axis [Fig.
\ref{Fig:M_Hmax}(b)], being also increasingly broader as $H_{max}$
increases up to fields of the order of 10$\times$ 10$^4$ Oe.
Another interesting observation is that the differences in
decreasing-field branches of the loops obtained after a FC
procedure for different $H_{max}$ are less significant than those
obtained after a ZFC procedure, while in increasing-field
branches, the differences are more significant after FC than after
ZFC procedure. Similar results are found for ferrihydrite
nanoparticles.

The horizontal shift of the hysteresis loops in the cooling field
direction is similar to that previously found in
ferritin,\cite{Berkowitz_prb},
ferrihydrite\cite{Seehra_ferrihydrite_doped} and other magnetic
nanoparticles.\cite{Oscar_JNN}
In the case of AF NiO nanoparticles, this loop shift was
associated to surface anisotropy and multisublattice states, with
the latter being associated to a variety of reversal
paths.\cite{Berkowitz_jmmmAF}
Surface anisotropy arises due to the breaking of the crystal-field
symmetry at the boundary of the nanoparticle. Two models have been
considered: one where the easy axis is transverse to the boundary
and another where the local easy axis depends on the site
``defect'' (N\'{e}el surface anisotropy
model).\cite{Kachkachi_anis_trans_Neel}
The loop shift in AF nanoparticles is also interpreted in terms of
an exchange bias between antiferromagnetic and uncompensated
moments, although no transition temperature is crossed but rather
a blocking temperature.

\begin{figure}[htb!]
\begin{center}\includegraphics[width=0.95\columnwidth]{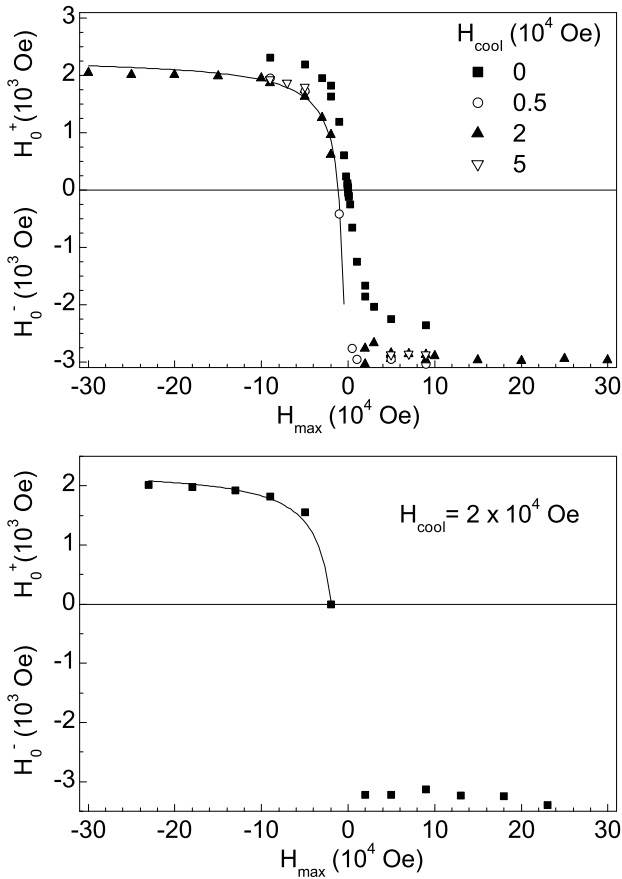} 
\end{center}
\caption{\label{Fig:H0_Hmax} Fields at which magnetization crosses
zero ($H_0^-$ and $H_0^+$) for different cooling fields $H_{cool}$
as a function of the maximum field $H_{max}$, for ferritin (a) and
ferrihydrite nanoparticles (b). Lines represent fit to Eq.
(\ref{H0_Hmax}).}
\end{figure}

\begin{figure}[htb!]
\begin{center}\includegraphics[width=0.9\columnwidth]{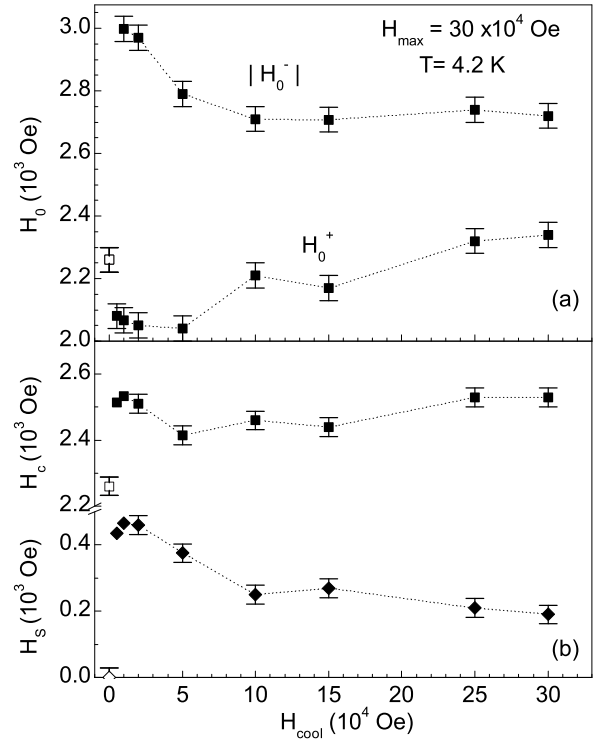}
\end{center}
\caption{\label{Fig:H0_Hcool}Dependence of the fields at which
magnetization crosses zero ($H_0^-$ and $H_0^+$) (a), $H_C$ and
$H_S$ (b) on the cooling field $H_{cool}$ for a maximum field
$H_{max}=30 \times 10^4$ Oe, for ferritin. The ZFC values are
shown in open symbols (which in the case of $H_0^-$ and $H_0^+$
corresponds to the same value). Lines connecting the FC values are
eye-guides.}
\end{figure}

The effect of $H_{max}$ and $H_{cool}$ on the field values at
which magnetization crosses zero in the decreasing and increasing field
branches of the hysteresis loop (termed $H_0^-$ and $H_0^+$, respectively) can be
observed in Fig. \ref{Fig:H0_Hmax}. For $H_{cool}\neq 0$, $H_0^-$
has a smaller variation with $H_{max}$ compared to that of
$H_0^+$, while for $H_{cool}=0$, $H_0^-$ and $H_0^+$ have
symmetric variations. 
>From Fig. \ref{Fig:H0_Hmax}, it is also clear that the effect of
$H_{max}$ on $H_0^+$ is more important than the effect of
$H_{cool}$.
In Fig. \ref{Fig:H0_Hcool}, we show the dependence of $H_0^+$,
$H_0^-$, the coercive field $H_C=(H_0^+ - H_0^-)/2$ and the loop
shift $H_S=-(H_0^+ + H_0^-)/2$ on the cooling field $H_{cool}$ for
the highest maximum applied field $H_{max}=30 \times 10^4$ Oe.
With increasing $H_{cool}$, $H_0^+$ increases approaching the ZFC
value for H$_{cool}\gtrsim 10^{4}$ Oe, whereas $|H_0^-|$ values
slightly decrease and are always higher in modulus than the value
for ZFC. Interestingly, the larger departures of $H_0^-$ and
$H_0^+$ from the ZFC value occur for lower $H_{cool}$. As a
result, $H_C$ is almost independent on $H_{cool}$, being higher
than the ZFC value, while the loop shift $H_S$ has a small
decrease with $H_{cool}$.
\subsection{Effect of $H_{max}$ and $H_{cool}$ in $M(t)$}

To have a better insight on the changes occurring near $H_0^-$ and
$H_0^+$ induced by the application of different $H_{max}$ and
$H_{cool}$, we have performed measurements of $M(t)$ near zero
field, after ZFC and FC under $H_{cool}=0.5 \times 10^4$ Oe as
described in Sec. \ref{Sec:Exp}. With these measurements, we aim
at demonstrating that the changes in the hysteresis loops are
related to changes in the energy barriers to magnetization
reversal induced by $H_{max}$ and $H_{cool}$.

\begin{figure}[htb!]
\begin{center}\includegraphics[width=0.9\columnwidth]{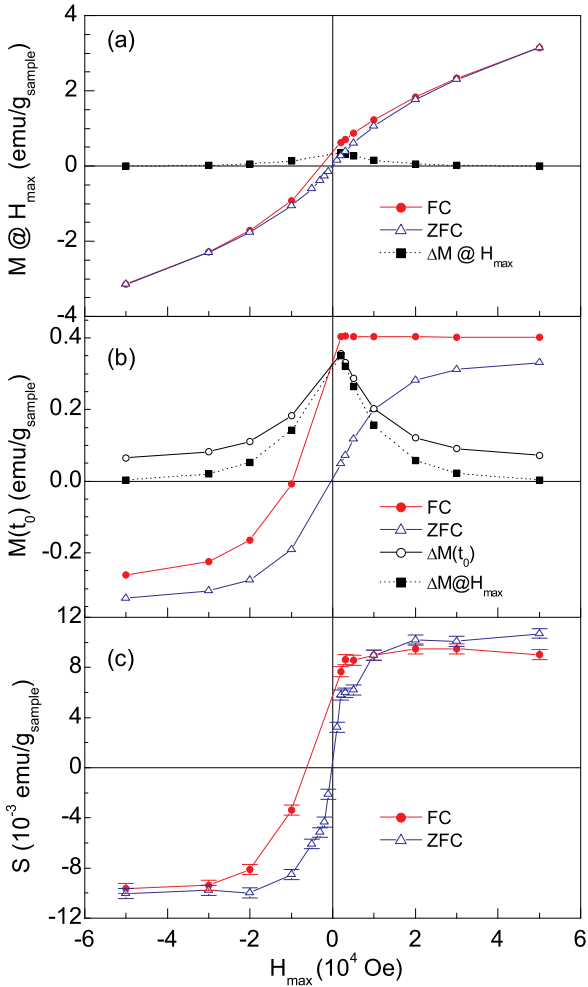}
\end{center}
\caption{\label{FigPara} (Colour online)(a) Magnetization measured at
$H_{max}$($-H_{max}$) before removing the field down to 50($-50$)
Oe ($M@H_{max}$) obtained after ZFC and FC with $H_{cool}=0.5
\times 10^4$ Oe, and difference between $M@H_{max}$ after FC and
ZFC $\Delta M@H_{max}$; (b) first value of the remanent
magnetization measured at 50($-50$) Oe [$M(t_0)$] obtained after
ZFC and FC with $H_{cool}=0.5 \times 10^4$ Oe and difference
between the FC and ZFC values; (c) magnetic viscosity $S$ measured
at 50($-50$) Oe after a previous $H_{max}$($-H_{max}$).
Measurements were performed in ferritin.}
\end{figure}

Before discussing the results of the magnetic relaxation
measurements presented in Fig. \ref{FigPara}(c), we will start by
analyzing the magnetic state of the samples attained after the
protocols previous to the relaxation measurements.
First, we plot in Fig. \ref{FigPara}(a) the values of
magnetization obtained at $H_{max}$ ($M@H_{max}$) after ZFC and FC
procedures. As it can be observed, the magnetization values
obtained after a FC are always higher than those attained after a
ZFC process.
Second, the magnetization values obtained right after the field is
decreased down to 50 Oe (-50 Oe) [$M(t_0)$] after FC and ZFC
procedures are plotted in Fig. \ref{FigPara}(b). Comparison with
results in the previous panel shows that the difference between
the ZFC and FC values of $M@H_{max}$ (full squares) is always
smaller than the differences between the ZFC and FC values of
$M(t_0)$ (open circles). In fact, at the highest applied
$H_{max}=5 \times 10^4$ Oe, FC and ZFC values of $M@H_{max}$ are
identical, while the corresponding values for $M(t_0)$ become
substantially different.
The differences after FC and ZFC procedures that appear in
$M(t_0)$ after decreasing the field down to $\pm$50 Oe are an
indication of the different energy barriers that each particle
magnetic moment has been able to cross. The higher $M(t_0)$ values
measured after FC hint at the appearance of energy barriers
imprinted after FC that are higher than after the ZFC process.
Moreover, the fact that $M@H_{max}$ values become the same after
measuring at the highest $H_{max}=5 \times 10^4$ Oe [see Fig.
\ref{FigPara}(a)] indicates that the differences in $M(t_0)$
appearing after both procedures cannot be attributed to an
increase of the net magnetic moment of the individual particles
induced at $H_{max}$ after the FC process.
Finally, it is also interesting to note the constancy of $M(t_0)$
observed for positive $H_{max}$ after FC [filled circles in Fig.
\ref{FigPara}(b)], which shows that the fraction of magnetic
moments that reverse after reduction of the field is almost
independent of the maximum applied field and reinforces the two
points commented previously. In contrast, when $H_{max}$ is
applied in a direction opposite to $H_{cool}$ [points with
negative abscissas in panels (a) and (b) of Fig. \ref{FigPara}],
there is a progressive increase of $M(t_0)$ with increasing
$H_{max}$ for both ZFC and FC procedures while $M@H_{max}$ values
are essentially the same in the two cases.
Again, this shows that the changes in $M(t_0)$ cannot be
attributed to an increase of the uncompensated moment of the
particles but rather to the fact that, when arriving near zero
field, different fractions of magnetic moments are able to cross
the energy barriers at $t_0$ depending on $H_{max}$ and
$H_{cool}$, an indication that the effective energy barriers felt
by the particles near zero field are modified by $H_{max}$. In a
simple picture, a negative $H_{max}$ has the effect of erasing the
barriers imprinted by the positive $H_{cool}$, with the system
being closer to the ZFC configuration as the intensity of
$H_{max}$ increases.

In what follows, we will analyze the results of the relaxation
measurements following the two above mentioned protocols. As
previously found in ferritin\cite{dMdH_prb, Mamiya_prl}, $M(t)$
displays a quasi-linear dependence on $\ln(t)$ at intermediate
times within the studied $t$ range (up to 1000 s) and can be
fitted to the following expression:
\begin{equation}\label{MrLinear} M(t)\approx M_0'-S\ln(t) \ ,
\end{equation}
where $M_0'$ is related to the initial magnetization and $S$ is
the so-called coefficient of magnetic viscosity. Eq.
(\ref{MrLinear}) is particularly useful in situations where
$\tau_0$ is not known.\cite{St_Pierre_prb} This equation can be
derived from a general expression for the time dependence of the
magnetization of an ensemble of nanoparticles with distribution of
energy barriers $f(E)$, after field removal
\begin{equation}\label{Mt}
M(t)=\int_0^\infty M_0(E)\exp(-t/\tau(E))f(E)dE\ ,
\end{equation}
where $M_0(E)$ is the initial magnetization of a particle with
energy barrier $E$.
It can also be shown that $S$ is proportional to $Ef(E)$ and,
therefore, $S$ is an appropriate quantity to observe changes in
the energy barriers.\cite{St_Pierre_prb, Jonsson_jmmm} However,
the direction of this change is not directly given by $S$ since,
in principle, $f(E)$ is a non monotonous function.
A distribution of energy barriers results directly from a
distribution of volumes, according to the relation
$f(E)=g(V)(dV/dE)$, where $g(V)$ is the volume distribution. Other
sources of a distribution of energy barriers are a distribution of
shapes, the existence of nanoparticles with the same size but with
different degrees of crystallinity, different oxygen and water
content.

This equation assumes that the magnetization decay of a
nanoparticle ensemble is due to the switching of the nanoparticles
magnetic moments as a consequence of energy barrier crossing when,
for a given $T$ and $H$, the Arrhenius relaxation time $\tau$ is
of the order of the measurement time $\tau_m$.

The $H_{max}$ dependence of the viscosity coefficient $S$ as
obtained from fits of the linear part of the relaxation curves to
Eq. \ref{MrLinear} is reported in Fig. \ref{FigPara}(c).
We observe that for positive $H_{max}$ and for the FC case, $S$
remains essentially constant with increasing $H_{max}$. This means
that $H_{cool}$ imprints energy barriers for reversal into the FC
direction that are not substantially changed by a positive applied
$H_{max}$. However, in the ZFC case, the relaxation rate increases
with $H_{max}$, showing that $H_{max}$ changes the energy barriers
in this case. For negative $H_{max}$ (applied contrary to the
$H_{cool}$ direction), however, the energy barriers are shifted.
In summary, the general behavior of $S$ is similar to that of
$M(t_0)$, reinforcing the interpretation of the effects of
$H_{cool}$ and $H_{max}$ in terms of energy barriers. Since in
ferritin \cite{FLuis_prb_ferritin,Mamiya_epl} and in the
ferrihydrite nanoparticles here studied \cite{NJO_dist}
interparticle interactions are negligible, the proposed changes in
the energy barriers are most probably associated to intra-particle
phenomena, as discussed below.

\subsection{Effect of $H_{max}$ on the energy barriers}

\begin{figure}[htb!]
\begin{center}\includegraphics[width=0.9\columnwidth]{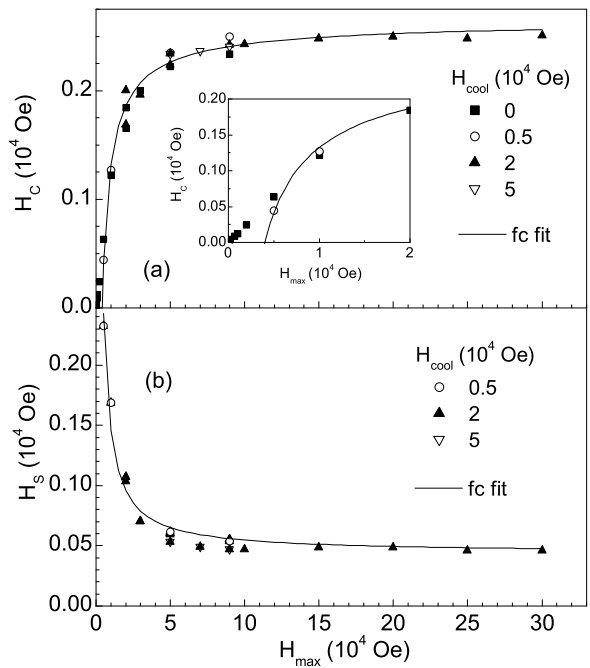}    
\end{center}
\caption{\label{Fig:H_C_Hmax} Dependence of the coercive field
$H_C$ (a) and loop shift $H_S$ (b) on the maximum field $H_{max}$
in ferritin, for different cooling fields $H_{cool}$. Lines in
panels (a) and (b) represent fits to Eq. (\ref{HC_Hmax}) and Eq.
(\ref{HE_Hmax}), respectively. Inset shows zoom over the low
$H_{max}$ region.}
\end{figure}

In Fig. \ref{Fig:H_C_Hmax}, we present the dependence of $H_C$ and
$H_S$ with $H_{max}$ for hysteresis loops measured after ZFC
(squares) and after cooling in different different $H_{cool}$. The
first point to notice is that, in the FC case, the loop shift
$H_S$ first rapidly decreases for low $H_{max}$ but, for fields
higher than $10\times10^4$ Oe, it saturates to a value around 500
Oe even for extraordinarily high values of $H_{max}= 30\times10^4$
Oe. This behavior is somewhat unexpected since, for the usually
observed loop shifts due to minor loops, the shift tends to zero
for sufficiently high fields.

Secondly, the constancy of $H_C$ obtained at high fields in the FC
case is an indication of the existence of high energy barriers
imprinted by $H_{cool}$ that cannot be surmounted even by applying
a $H_{max}$ of $30\times10^4$ Oe field in the direction opposite
to $H_{cool}$. Both observations (together with the $H_0^+$
variation already presented in Fig. \ref{Fig:H0_Hmax}) indicate an
initial evolution of the minor loops due to crossing of smallest
energy barriers. After this first stage, the variations are
smoothed by the higher energy barriers imprinted by $H_{cool}$.
According to the behavior of $H_0^+$, $H_0^-$ and $M(t_0)$, it is
clear that $H_{cool}$ increases the energy barriers in the field
direction ($E_0^-$) and decreases barriers in the opposite
direction ($E_0^+$). Since the highest barriers are not overcome,
the symmetric situation $E_0^+=E_0^-$ cannot be recovered and thus
$H_S$ is always different from zero.

A quantitative description of these phenomena can be given within
the framework of the uniform rotation model in terms of the
influence of $H_{max}$ on the energy barriers near zero field
$E_0$. First, let us notice that the dependence of $H_C$ on
$H_{max}$ is similar to the dependence of $H_C$ on $V$ usually
found in nanoparticle systems (see Eq.
\ref{H_CT}).\cite{Rivas_HC_random} Taking into account this
resemblance, and with the aim to propose an expression for
$H_C(H_{max})$ which properly describes the measured data, we will
assume that $H_{max}$ influences the zero field energy barriers in
a way similar to the particle volume (Eq. \ref{EKV})
\begin{equation}
\label{gamma} E_0\propto H_{max}^\gamma
\end{equation}
where $\gamma$ is a power law exponent that controls the way $H_C$
approaches its limiting value for high $H_{max}$, such that a higher $\gamma$ is associated to a faster approach to saturation.  
This power law dependence condensates different possible
mechanisms for the influence of $H_{max}$ on intrinsic energy
barriers: either a change in the exponent $p$ in Eq. (\ref{EKV})
or an irreversible increase of the anisotropy constant $K$. The
first possibility seems to be ruled out since, as reported in the
previous section, $H_{max}$ does not seem to affect the net
magnetic moment of the particles. Therefore, the influence of
$H_{max}$ can be thought mostly as an effect on $K$ associated to
an increase of the local (intra-particles) energy barriers. Since
these are macroscopically average measurements, it is difficult to
access the ``microscopic'' origin for this effect on $K$. Anyway,
this can be understood considering that the system has multiple
configurations with associated energies such that $H_{max}$ and
$H_{cool}$ selects or imprints a set of these configurations
restricting the relaxation of the moments.

Based on Eq. (\ref{HC_E}) the relation between $H_0^+$ and
$H_{max}$ can be expressed as
\begin{equation}
\label{H0_Hmax}
H_0^+(H_{max})=H_0^+(\infty)\left[1-\left(\frac{H_B^+}{H_{max}}\right)^{\beta_+}\right]
\end{equation}
where $\beta_+=\alpha / \gamma$, $H_0^+(\infty)$ is redefined as
$H_0^+$ for infinite $H_{max}$ and $H_B^+$ is defined as the field
at which $H_{0}^{+}$ is zero. The relation between $E_0$ and
$H_{B}^{+}$ is that expressed in Eq. (\ref{gamma}) since
$H_{B}^{+}$ is a particular case of an $H_{max}$. When comparing
to experimental data, care must be taken and the FC and ZFC cases
must be distinguished. In the FC case, $H_{0}^{+}$ can be negative
and $H_{B}^{+}$ is the field at which $H_{0}^{+}$ crosses zero
(Fig. \ref{Fig:H0_Hmax}).
In the ZFC case, $H_{B}^{+}$ is defined as the field at which the
behavior of $H_{0}^{+}$ at high $H_{max}$ extrapolates to zero,
since in practice the experimental $H_{0}^{+}$ values are not zero
at $H_{max}\leqq H_{B}^{+}$.

Eq. (\ref{H0_Hmax}) can successfully describe the $H_0^+(H_{max})$
data shown in Fig. \ref{Fig:H0_Hmax}, with
$H_0^+(\infty)=0.23\times10^4$ Oe, $H_B^+=-1.1\times10^4$ Oe and
$\beta_+=0.8$ for the FC data. In the case of ferrihydrite
$H_0^+(\infty)=0.20\times10^4$ Oe, $H_B^+=-2.0\times10^4$ Oe,
$\beta_+=1.5$. Considering $\alpha=4/3$, $\gamma\approx1$ in
ferritin and $\gamma\approx2$ in ferrihydrite, i. e. approximately
a linear and quadratic relation between $E_0$ and $H_{max}$. The
differences here found for $\gamma$ are associated to the fact
that in ferrihydrite a smaller $H_{max}$ is enough to approach
$H_0^+$ to saturation. Again, the ``microscopic'' origin for this
mechanism is not clear.

For ZFC data, both $H_0^+(H_{max})$ and $H_0^-(H_{max})$ are well
described by Eq. (\ref{H0_Hmax}) for $|H_{max}|>|H_{B}^{+}|$,
while for FC $H_0^-(H_{max})$ is approximately constant.
Accordingly, the generalization to the $H_{max}$ dependence of the
coercive field for ZFC and FC procedure ($H_{CZFC}$ and $H_{CFC}$,
respectively) and of $H_S$ is straightforward
\begin{eqnarray}
\label{HC_Hmax}
H_{CZFC}(H_{max})=H_0^+(\infty)\left[1-\left(\frac{H_B^+}{H_{max}}\right)^{\beta_+}\right]
\nonumber \\
H_{CFC}(H_{max})=-\frac{1}{2}H_0^-
+\frac{1}{2}H_0^+(\infty)\left[1-\left(\frac{H_B^+}{H_{max}}\right)^{\beta_+}\right]
\end{eqnarray}
and
\begin{equation}
\label{HE_Hmax} H_S(H_{max})=-\frac{1}{2}H_0^-
-\frac{1}{2}H_0^+(\infty)\left[1-\left(\frac{H_B^+}{H_{max}}\right)^{\beta_+}\right]
\end{equation}
where $H_0^-$ is constant.

As expected from the agreement between the $H_0^+(H_{max})$ data
and Eq. (\ref{H0_Hmax}), Eq. (\ref{HC_Hmax}) can also be
successfully used to describe the $H_{CZFC}(H_{max})$ and
$H_{CFC}(H_{max})$ data 
in the $0.5\times 10^4 < H_{max} < 30\times 10^4$ Oe range, 
as shown in Fig. \ref{Fig:H_C_Hmax}(a), with $H_0^+=
0.24\times10^4$ Oe and $H_B^+=0.9\times10^4$ Oe for the ZFC data,
and with the previous $H_0^+(H_{max})$ parameters and
$H_0^-=-0.297\times10^4$ Oe for the FC data.
For low fields ($H_{max}<0.5\times 10^4$ Oe) the fit deviates from
the experimental data, while the fit considering only the high
field data extrapolates to $H_C=0$. At $H_B^+$, the experimental
$H_C$ is still about 18\% of its saturation value, approaching
zero for $H_{max}=0$. The differences between the $H_{CZFC}$ and
$H_{CFC}$ data are probably of the order of data error and, thus,
the differences between ZFC and FC fitted parameters are also
within the error bars.

The value of $H_C$ at $H_{max}=5\times 10^4$ Oe and 3.2 and 4.2 K
is of the order of that previously found for ferritin at 5 K
($\sim 1700$ Oe).\cite{Berkowitz_prb} The slightly higher value
that was found (2200 Oe) is probably due to the lower temperature.
Other factors affecting $H_C$ that may contribute to this
difference are the field sweeping rate and characteristic time of
measurement.

As in the case of $H_C$, the dependence of $H_S$ with $H_{max}$
can be in fact described by Eq. (\ref{HE_Hmax}), with the
parameters obtained for $H_0^+$ and a constant
$H_0^-=-0.297\times10^4$ Oe, which gives an extrapolated
$H_S(\infty)=335$ Oe.

\section{Conclusions}

Coercivity and loop shifts in nanoparticles are dynamical
phenomena, which depend on temperature, characteristic time of
measurement and number of cycles, for instance. In ferritin, we have
shown that coercivity and loop shifts depend also on the cooling
field and on the maximum field used, for fields higher than those
normally used. The dependence of coercivity and loop shifts with
the maximum field can be described in terms of changes in the
anisotropy energy barrier near zero field induced by the maximum
field, and quantitatively described by a modified N\'{e}el-Brown
model here proposed. Qualitatively, field cooling imprints energy
barriers, such that the energy barriers near zero in the
descending and ascending branches of the magnetization cycle are
higher and lower than in the ZFC case, respectively. This
difference is attenuated (but not erased) by increasing the
maximum field in the opposite direction of the cooling field.
Accordingly, the loop shift decreases with the maximum field but
it is not zero up to the highest field used ($30\times 10^4$ Oe),
showing that the barriers imprinted by field cooling cannot be
overcome by these high fields.

The experimental observations and subsequent analysis presented in
this article have evidenced the imprinting of high energy barriers
through an effective anisotropy induced by the applied protocols.
This gives rise to the high irreversibility and minor loop effects
similar to those observed in spin-glasses and diluted
antiferromagnets, where this phenomenology is ascribed to dilution
and the antiferromagnetic character of the interactions and not to
frustration.\cite{LevySG,LabartaSG}
%

\begin{acknowledgments}
The authors acknowledge V. de Zea Bermudez for the synthesis of
the organic-inorganic hybrid containing ferrihydrite nanoparticles
and R. Boursier for his help with the high field setup in
Grenoble. Part of this work has been supported by EuroMagNET II
under the EU contract No. 228043. We acknowledge IFIMUP for the
possibility of performing the magnetization time dependence
measurements. The financial support from FCT (Grant No.
PTDC/FIS/105416/2008) is gratefully recognized. The
Aveiro-Zaragoza collaboration has been supported by the Integrated
Spanish-Portuguese Action under the Grant No. PT2009-0131. The
work in Zaragoza has been supported by the research Grants No.
MAT2007-61621 and CONSOLIDER CSD2007-00010 from the Ministry of
Education. \`{O}. I. and A. L. acknowledge funding of the Spanish
MICINN through Grant No. MAT2009-08667 and No. CSD2006-00012, and
Catalan DIUE through project No. 2009SGR856. N. J. O. S.
acknowledges CSIC for a I3P contract and FCT for Ciencia 2008
program, A. U. acknowledges the financial support provided by
contract from the EC NoE `MAGMANET' and R. B. acknowledges
ICMA-CSIC for a JAE-predoc grant.
\end{acknowledgments}

\bibliography{bib_NJOSilva}

\end{document}